\documentclass[a4paper,11pt]{article}
\pdfoutput=1 % if your are submitting a pdflatex (i.e. if you have
\usepackage{jheppub} % for details on the use of the package, please
\usepackage{accents}

\usepackage[T1]{fontenc} % if needed

\title{Electromagnetic fields and charges in expanding universes}
\author[a]{Luciano Combi,}
\author[a,b]{Gustavo E. Romero}
\affiliation[a]{Instituto Argentino de Radioastronom\'ia (CCT-La Plata, CONICET; CICPBA), C.C. No. 5, 1894, Villa Elisa, Argentina}
\affiliation[b]{Facultad de Ciencias Astron\'omicas y Geof\'isicas, Universidad Nacional de La Plata, Paseo del Bosque s/n, 1900 La Plata, Buenos Aires, Argentina.}
\emailAdd{lcombi@iar.unlp.edu.ar}

\abstract{We analyze the properties of the electric and magnetic fields in different reference frames within a cosmological background space-time. First, we investigate the conformal properties of the electromagnetic fields and charge currents, discussing how the spatial curvature of the universe affects the field on different scales. Then, we analyze the effects of the expansion of the universe on local electromagnetic sources using Fermi coordinates. In particular, we investigate the energy balance and Poynting flux in this locally defined reference frame. We show that a charge following the Hubble flow in an accelerated FLRW universe is accelerated as seen by the local inertial frame, leading to non-null radiation.}

\begin{document} 
\maketitle
\flushbottom

\section{\label{sec:level1} Introduction}

% Addition + Change of order
The frame dependency of the electromagnetic field has been widely discussed in the literature both in classical and quantum physics. In curved space-time, an apparent paradox appears when we try to join the concept of radiation and the equivalence principle. As was clarified by Rohrlich \cite{rohrlich1963principle} and Boulware \cite{boulware1980radiation}, the radiation of a charged particle is not a covariant concept: if a charge is static in a gravitational field, an observer in a free-falling frame measures radiation while an observer comoving with the charge does not. 

% Addition
Radiation of charged particles has been analyzed in different scenarios and in the context of different space-times (see for instance Refs. \cite{Kuchar:2013bla}, \cite{Higuchi:1996aj} and \cite{Trzetrzelewski:2015rga}) but has not been treated in a cosmological background. In the standard model of cosmology, the universe on large scales is described by the homogenous and isotropic Friedmann-Lemaitre-Roberston-Walker (FLRW) space-time. The metric that represents this model is locally conformal to the Minkowski metric. The analysis of electromagnetic fields in a cosmological background is then simplified because Maxwell's equations are conformal invariant: well-known results in flat space-time can be mapped to an expanding universe. For this reason, as measured in the frame comoving with the cosmological fluid, the electric and magnetic fields always decay adiabatically as the inverse square of the scale factor, $E,B \sim a(t)^{-2} \accentset{\circ}{E},\accentset{\circ}{B}$ where $\accentset{\circ}{E},\accentset{\circ}{B}$ are defined in Minkowski space-time. The electric and magnetic fields are, however, frame-dependent properties of the electromagnetic field. The adiabatic decay will only occur for a frame comoving with the cosmic dust in a spatially flat universe. Changing either the spatial curvature of the universe or the reference frame implies that the fields will be more complex than in Minkowski space-time. 

%Addition
On the other hand, in non-asymptotically flat space-times, radiation as a far-zone field is non-trivial and not yet well-understood  (see Refs. \cite{Ashtekar:2014zfa} and \cite{Ashtekar:2015lla}). A way to analyze radiation, and energy balance in general, is to consider quasi-local quantities, where a congruence of observers must be specified to compute well-defined physical quantities. In order to do this, a very rigorous analysis is needed. This issue is ultimately important to understand the Unruh effect and the role of quantum fields in cosmology (see e.g. Refs. \cite{benevides2018b} and \cite{Matsui:2018iez}). The  situation of charges in cosmology has been recently discussed in the framework of de Sitter space-time \cite{akhmedov2010classical} for some specific cases, but no comprehensive treatment is available in the literature for a general FLRW metric.

The main goal of this work is to analyze the frame dependency of the electric and magnetic fields of local sources within an expanding universes. In Sec. 2 we present the covariant formalism used to analyze the electric and magnetic fields in a given reference frame for a general space-time background and we analyze the conformal properties of the EM field. In Sec 3. we apply this to the local observer, constructed using Fermi normal coordinates. Then, we revisit the charged particle field as described from this frame. In particular, we investigate whether if a charge accelerating with the universe produces radiation and the differences that appear with the usual cosmic frame.

Notation: Throughout the paper we use Latin letters $a= (0),(1),(2),(3)$ for Lorentz indices and Greek letters $\mu=0,1,2,3$ for holonomous coordinates. Bold letters as $\mathbf{V}$ are four-vectors and arrow letters as $\vec{U}$ are three-vectors, i.e. projection to a spatial base. 

\section{Electromagnetic fields in conformally flat space-times}

\subsection{Frames in space-time}

In a given space-time, a reference system frame is represented as a congruence characterized by a tetrad field $\mathbf{e}_{a}$ (and the cotetrad $\mathbf{e}^{a}$), with a tetravelocity $\mathbf{e}_{(0)}=\mathbf{u}$ and spatial triads $\mathbf{e}_{(i)}$ for each path in the congruence. The components of the tetrad field obey the orthonormal relation given by:
\begin{equation}
g_{\mu \nu} = e_{\mu}^{a} e_{\nu}^{b} \eta_{ab}.
\label{eq:obs}
\end{equation} 

Any vector field $\mathbf{V}$ can be expanded in a given frame as $\mathbf{V}:= V^{a}\mathbf{e}_{a}$, where $V^{a}=V^{\mu} e^{a}_{\mu}$. We use $\vec{V}:= V^{(i)} \mathbf{e}_{(i)}$ for the spatial part. At each point of the manifold, different frames are related by local Lorentz transformations. The kinematic features of the frame are encoded in the Ricci rotation coefficients (see Refs. \cite{Felice:2010cra} and \cite{Costa_Natario_2014}):
\begin{equation}
e^{\nu}_{b} \nabla_{\nu} e_{c}^{\mu} = \Gamma^{a}_{bc} e_{c}^{\mu}.
\end{equation} 

From here, we can obtain the acceleration of the frame, given by $a^{(i)}:= \Gamma^{(i)}_{(0)(0)}= e^{(i)}_{\mu} u^{\nu} \nabla_{\nu} u^{\mu}$; hence, only frames in free fall are not accelerated. Other characteristics of the congruence follows from:
\begin{equation}
\Gamma^{(0)}_{(i)(j)}= \sigma_{(i)(j)} + \frac{1}{3} \Theta \delta_{(i)(j)} + \omega_{(i)(j)},
\end{equation}
where $\sigma_{(i)(j)}$, $\Theta$, and $\omega_{(i)(j)}$ are the projected components of the shear, the expansion, and the vorticity, respectively. It is also useful to define:
\begin{equation}
h_{\mu \nu}=g_{\mu \nu} + u_{\mu} u_{\nu}= e^{(i)}_{\mu}e^{(j)}_{\nu} \delta_{(i)(j)},
\end{equation}
as the projector to the congruence rest-space, and the spatial derivative of a tensor as:
\begin{equation}
D_{\mu} X^{\sigma}:= h_{\mu}^{\alpha} h^{\sigma}_{\beta} \nabla_{\alpha} X^{\beta}.
\end{equation}

Given a frame, we introduce the connection for spatial vectors $\vec{\nabla}$, defined in tetrad components as:
\begin{equation}
\vec{\nabla}_{(i)} X^{(j)}= X^{(j)}_{,(i)} + \Gamma^{(j)}_{(i)(k)} X^{(k)} \equiv e_{(i)}^{\mu}e^{(j)}_{\nu} D_{\mu} X^{\nu}.
\end{equation}

Defining the totally antisymmetric tensor $\eta_{\rho \mu \nu \sigma} :=\sqrt{-g} [\rho \mu \nu \sigma]$, we have the spatial Levi-Civita tensor as $\epsilon_{\mu \nu \sigma}:= \eta_{\rho \mu \nu \sigma} u^{\rho}$. With this, we define the spatial curl as:

\begin{equation}
( \vec{\nabla} \times \vec{V})_{(i)}:= \epsilon_{\mu\nu\sigma} e_{(i)}^{\mu} D^{\nu}V^{\sigma}= \epsilon_{(i)(j)(k)} \vec{\nabla}^{(j)} V^{(k)},
\end{equation}
and the divergence as:
\begin{equation}
\vec{\nabla} \cdot \vec{V}:=  D_{\mu} V^{\mu}.
\end{equation}

\subsection{Covariant electromagnetic field equations}

The electromagnetic field is represented by the antisymmetric Faraday's tensor $F_{\mu \nu}$. Maxwell's equations on a fixed background space-time are given by:

\begin{equation}
\nabla_{\mu} F^{\mu \nu}= 4 \pi j^{\nu},
\end{equation}
\begin{equation}
\nabla_{[\mu} F_{\rho \sigma]} =0.
\end{equation}

The energy-momentum tensor of the electromagnetic field is:
\begin{equation}
T^{\mu \nu}= \frac{1}{4 \pi } \Big[ F^{\mu \alpha} F^{\nu}_{\; \; \; \alpha} - \frac{1}{4} F^{\alpha \beta} F_{\alpha \beta} \Big],
\end{equation}
holding
\begin{equation}
\nabla_{\mu} T^{\mu \nu}= j_{\mu} F^{\mu \nu},
\label{eq: conserv}
\end{equation}
when the field equations are satisfied. Although Maxwell's equations are Lorentz invariant, and also general invariant when combined with Einstein's equations, the electromagnetic field has different properties in different reference frames. Relative to a congruence, the Faraday's tensor can be decomposed as \cite{de1992relativity}:

\begin{equation}
F_{\mu \nu}= 2 u_{[\mu} E_{\nu]}+ \epsilon_{\mu \nu \rho} B^{\rho},
\end{equation}
where the electric and magnetic fields covectors are defined, respectively, as:
\begin{equation}
E_{\mu}:= F_{\mu \nu} u^{\nu}, \quad B_{\mu}:= \frac{1}{2} \epsilon_{\mu \nu \rho} F^{\rho \nu}\equiv \star F_{\mu \nu} u^{\nu},
\end{equation}
where $ \star$ is the Hodge product of a form. The projected components of these fields over the tetrad field given by $E^{a}=e^{a}_{\mu} E^{\mu}$ and $B^{a}=e^{a}_{\mu} B^{\mu}$ shows that $E^{(0)}=B^{(0)}=0$, i.e. the electric and magnetic fields are space-like vectors with respect to the frame. The electromagnetic current $j^{\mu}$ can be written as:
\begin{equation}
j^{\mu}= \rho_{e} u^{\mu} + J^{\mu},
\end{equation}
where $\rho_{e}:=j^{(0)}$ and $J^{\mu}:= e^{\mu}_{(i)} j^{(i)}$. Projecting Maxwell's equations on this frame gives two propagation equations \cite{Tsagas_2005}:
\begin{equation*}
\dot{E}_{(i)}= \sigma_{(i)(j)}E^{(j)}+ (\omega \times \vec{E})_{(i)} - \frac{2}{3}\Theta \vec{E}_{(i)} 
\end{equation*}
\begin{equation}
    + (\vec{a} \times \vec{E})_{(i)} + (\vec{\nabla}\times \vec{B})_{(i)} -  j_{(i)},
\end{equation}

\begin{equation}
\dot{B}_{(i)}=  \sigma_{(i)(j)}B^{(j)}+ (\omega \times \vec{B})_{(i)}- \frac{2}{3}\Theta \vec{B}_{(i)} -(\vec{a} \times \vec{B})_{(i)} - (\vec{\nabla}\times \vec{E})_{(i)},
\end{equation}
and two constrain equations:
\begin{equation}
\vec{\nabla} \cdot \vec{E} + 2 \vec{\omega}\cdot \vec{B}= \rho_{e}, \quad \vec{\nabla} \cdot \vec{B} - 2 \vec{\omega}\cdot \vec{E}= 0.
\end{equation}
%Addition
where we define $\vec{\omega}_{(i)}:=\frac{1}{2} \epsilon_{(i)(j)(k)} \omega^{(j)(k)}$. Note that in a Minkowski inertial frame, this set of equations reduces to the well-known three-vector form of Maxwell's equations. The evolution of the local charge density $\rho_e$ for a given congruence is given by
\begin{equation}
\dot{\rho}_e = -\Theta \rho_e -D^{\mu}J_{\mu} - a^{\mu}J_{\mu}.
\end{equation}

Since $j^{\mu}$ is a conserved vector for any $F_{\mu \nu}$, the total electric charge defined as $Q=\int_{\Sigma} J^{\mu} d^3 \Sigma_{\mu}$ in a space-like hypersurface $\Sigma$ is conserved, as required from the gauge invariance of the theory. We can decompose $T^{\mu \nu}$ in terms of a given frame $\mathbf{e}_a$ in the following way:
\begin{equation}
4 \pi T_{\mu \nu}= \frac{1}{2} (E^2+B^2) u_{\mu}u_{\nu} + \frac{1}{6}(E^2+B^2) h_{\mu \nu} + 2 \mathcal{Q}_{(\mu}u_{\nu)} +\pi_{\mu \nu},
\end{equation}
where $E^2=E_{\mu}E^{\mu}$, $B^2=B_{\mu}B^{\mu}$ are the magnitudes of the fields, $\mathcal{Q}_{\mu}:=\epsilon_{\mu \nu \rho} E^{\nu} B^{\rho}$ is the Poynting vector, and $\pi_{\mu \nu}$ the anisotropic stress defined as
\begin{equation}
\pi_{\mu \nu}:= \frac{1}{3} (E^2+B^2) h_{\mu \nu} -E_{\mu} E_{\nu} -B_{\mu}B_{\nu},
\end{equation}

In this formulae, we can interpret the electromagnetic field as an imperfect fluid with energy density $\rho= (E^2+B^2)/2$, isotropic pressure $p=\rho/3$, energy-flux $\mathcal{Q}^{\mu}$, and anisotropic stress $\pi_{\mu \nu}$. Projecting (\ref{eq: conserv}) over $\mathbf{u}$, we obtain the local energy conservation with respect to the congruence:
\begin{equation}
\dot{\rho}= -\frac{4}{3}\Theta \rho - D^{a} \mathcal{Q}_{a} - 2 a^{\mu} \mathcal{Q}_{\mu} - \sigma^{\nu \mu} \pi_{\mu \nu} + E_{\mu} J^{\mu}.
\label{eq: econserv}
\end{equation} 

Here, we also see that although the energy-momentum is Lorentz invariant, the energy density and energy transfer of the fields depends on the reference frame. From Maxwell's equations, Faraday's tensor $\mathbf{F}\equiv \: F_{\mu \nu} \mathbf{d}x^{\mu} \wedge \mathbf{d}x^{\nu}$ is a closed two-form, $\mathbf{dF}=0$, so we can write it as $\mathbf{F}=\mathbf{d}\mathbf{A}$, or in component notation:
\begin{equation}
F_{\mu \nu} \equiv \nabla_{\mu} A_{\nu} - \nabla_{\nu} A_{\mu}.
\end{equation}

The potential vector $A^{\mu}$ can be decomposed in a temporal and spatial in the orthonormal frame, $\mathbf{A}\equiv A^{(0)} \mathbf{u} +\vec{A}$. It is possible to show that the magnetic and electric fields formulas in terms of the potential still hold in the covariant description as:
\begin{equation}
\vec{B}= \vec{\nabla}\times \vec{A}, \quad \vec{E}= - \dot{\vec{A}} + \vec{\nabla}A^{(0)}.
\end{equation}

Note that the expression for the magnetic field is valid even if it has a non-null divergence, i.e. $\vec{E}\cdot \vec{\omega} \neq 0$ \footnote{The vector identity for the divergence and the curl that holds in Euclidean space does not hold in spatial vectors in curved space-time since the Levi-Civita tensor depends on the metric.}.

\subsection{Conformal invariance and frames}

Since $F^{\mu \nu}$ is antisymmetric and the Levi-Civita's connection is symmetric, Maxwell's equations are equivalent to
\begin{equation}
\partial_{\mu} \mathcal{F}^{\mu \nu}= 4 \pi \mathcal{J}^{\nu},
\label{eq: max1}
\end{equation}
\begin{equation}
\partial_{[\mu} F_{\rho \sigma]} =0,
\label{eq: max2}
\end{equation}
where $\mathcal{F}:= \sqrt{-g} F^{\mu \nu}$, $\mathcal{J}^{\nu}:= \sqrt{-g} j^{\nu}$. If we assume that $F_{\mu \nu}$ does not change under conformal transformations,
\begin{equation}
\accentset{\circ}{g}_{\mu \nu} \rightarrow  g_{\mu\nu}=\Omega^2(x)\: \accentset{\circ}{g}_{\mu \nu},
\label{eq: metconf}
\end{equation}
from equations (\ref{eq: max1}) and (\ref{eq: max2}) it can be shown that Maxwell's theory is conformally invariant\footnote{Note that the coordinates do not change in this scale transformation and that, $\sqrt{-g}=\Omega^4 \sqrt{-\overline{g}}$, $F^{\mu \nu}= \overline{F}^{\mu\nu} \Omega^{-4}$, and $j^{\mu}= \Omega^{-4} \overline{j}^{\mu}.$}. This means that Maxwell's equations in any space-time conformally related to Minkowski's will have the same well-known flat solutions, at least locally. Let us note, however, that (a) the solutions of  $F^{\mu \nu}$ will map exactly only if the conformal transformation is global (see Refs. \cite{Infeld_Schild_1945}, \cite{Infeld_Schild_1946}, and \cite{Barrow_Tsagas_2008}), (b) the electric and magnetic fields measured by inertial observers in a conformally flat space-time (CFS) will not coincide with their values in Minkowski space-time since these frame are not equivalent and (c) the highly symmetric case of a FRLW universe is incompatible with the anisotropy of the energy-momentum tensor of the EM field; one can either work in a perturbed FRLW metric, which is no longer a CFS  \cite{Tsagas_2014}, or analyze the fields in this fixed background. Electric and magnetic fields, in turn, will have nontrivial characteristics in the CFS as we shall see.

Let us analyse general features of the reference frame in a CFS. Consider a frame $\mathbf{e}_a$ in a conformally flat space-time. This tetrad field is orthonormal, $g_{\mu \nu} = e_{\mu}^{a} e_{\nu}^{b} \eta_{ab}$ with respect to the conformal metric $g_{\mu \nu}$. From \ref{eq:obs}, the vector basis $\accentset{\circ}{\mathbf{e}}_{a} =  \Omega(x) \: \mathbf{e}_{a} $ (and the cobasis $\accentset{\circ}{\mathbf{e}}^{\:a} = \mathbf{e}^{a} / \Omega(x)$) is a Minkowski-adapted tetrad field, i.e. 
\begin{equation}
\accentset{\circ}{e}^{\: \:a}_{\mu} \accentset{\circ}{e}^{\: \: b}_{\mu} \eta_{ab} = \eta_{\mu \nu}.
\end{equation} 

If $\accentset{\circ}{\mathbf{e}}_{a}$ is an inertial observer (i.e., a frame which assumes the form $\accentset{\circ}{e}^{\: \mu}_{\: a}=\delta^{\mu}_{a}$ in Cartesian coordinates), then the kinematic properties of $\mathbf{e}_{a}$ are determined by the conformal factor $\Omega(x)$ because $\mathbf{e}_{a}= \delta^{\mu}_{a}/\Omega(x) \partial_{\mu}$ . The acceleration of this frame is easily obtained as:
\begin{equation}
a_{(i)}= e_{(i)}^{\mu} e_{(0)}^{\nu} \nabla_{\nu} e_{\mu}^{(0)}= \frac{\partial_{i} \text{log}(\Omega)}{\Omega}.
\label{eq: accelframe}
\end{equation}

The congruence is also characterized with a non-null shear and expansion:
\begin{equation}
\sigma_{(i)(j)}= \delta_{(i)(j)} \frac{\partial_{t} \text{log}(\Omega)}{\Omega}, \quad \Theta= 3 \frac{\partial_{t} \text{log}(\Omega)}{\Omega}.
\label{eq: exp}
\end{equation}

We shall analyze first an only time-dependent conformal factor, as appears in flat FLRW universes, with line element:
\begin{equation}
ds^2= \Omega^2 (t) (-dt^2 + d^3x).
\end{equation}

In this case, the conformal observers are inertial, i.e. $a_{(i)}=0$, but the congruence has an isotropic expansion (see equation (\ref{eq: exp})). Later on, we shall discuss space curved universe and anisotropic scale factors. This means that the observer represented as $\mathbf{e}_a$ will measure a decay (grow) of the fields if the conformal factor grows (decays):
\begin{equation}
\vec{E}= \Omega^{-2} \accentset{\circ}{
\vec{E}}, \quad \vec{B}= \Omega^{-2} \accentset{\circ}{\vec{B}},\
\label{eq: emcosm}
\end{equation}
where $\accentset{\circ}{\vec{E}}$ and $\accentset{\circ}{\vec{B}}$ are the electric and magnetic fields measured by an inertial Minkowski observer. Most applications in cosmology are derived from the use of FLRW coordinates. The line element in conformal coordinates transforms to these coordinates according to
\begin{equation}
dt_{F}= \Omega(t)\: dt,
\end{equation}
where the metric assumes the well-know form:
\begin{equation}
ds^2= -dt_{F}^2 + \Omega (t_{F})^2 \: \delta_{ij} \: dx^i_{F} dx^j_F.
\label{eq: flrw}
\end{equation}

Stationary observers in these coordinates, that we call \textit{cosmic observers}, represented by $\mathbf{e}^F_{a}=\lbrace \partial_{t_{F}},  \partial_{i_{F}} /\Omega \rbrace$, are inertial, and $t_F$ is the cosmic time; this frame is comoving with the cosmic fluid, also known as the \textit{Hubble flow}. It can be shown that these stationary frames coincide with the conformal observers, $\mathbf{e}^F_{a}\equiv \mathbf{e}_{a}$. Thus inertial observers in a spatially flat comoving with the Hubble flow measure electric and magnetic fields of the form (\ref{eq: emcosm}).

\subsection{Charges in conformally flat space-times}

%Addition%%%%%%%%%%%%%%%%%%%%%%%%%%%%%%%
Charges in free-fall can produce radiation, as it is well-known from a charged particle orbiting around a massive body. As we shall show below, this could also occur in a conformally flat space-time.
%%%%%%%%%%%%%%%%%%%%%%%%%%%%%%%%%%%%%%%%%
Solutions of Maxwell's equations in a spatially-flat FLRW space-time can be obtained directly from flat space-time solutions. Charge currents, however, are not exactly transformed since geodesics in flat space-time are not necessarily geodesics in the conformal space-time, except when the geodesic is null \cite{Wald_2010}. The mapping of charge currents can be written as $(F_{\mu \nu}, Q,\mathbf{v}) \rightarrow (F_{\mu \nu},Q,\accentset{\circ}{\mathbf{v}})$, where the velocities do not share the same kinematic state.

Let us consider a charge with a tetra-velocity $\mathbf{v}$ following a path $\gamma$ in a FLRW space-time. If the charge is geodesic, then we have $\nabla_{\mathbf{v}} \mathbf{v} = 0$. The solution of Maxwell's equations with this source is given by a flat space-time solution of a charge with a transformed tetra-velocity $\accentset{\circ}{\mathbf{v}}= \mathbf{v} \Omega$ which is not geodesic in general. In the particular case where the charge is comoving with the Hubble flow, 
\begin{equation}
\mathbf{v}= \partial_{t_{F}}\equiv\partial_{t}/\Omega \rightarrow \accentset{\circ}{\mathbf{v}}=\partial_t.
\end{equation}

The corresponding charge in the conformally transformed space is then geodesic and the solution is the Coulomb field, that is mapped to the FLRW space-time as in (\ref{eq: emcosm}). If the charge, however, is geodesic but has a non-null peculiar velocity, its tetra-velocity is
\begin{equation}
\mathbf{v}= \gamma ( \mathbf{e}^F_{(0)} + v^{(i)}\mathbf{e}^F_{(i)}),
\end{equation}
where $\gamma:= 1/\sqrt{1-v^2}$, and $v^2=\delta_{(i)(j)} v^{(i)}v^{(j)}$. It is well known that the spatial components decay as $v \sim 1/\Omega$, i.e. geodesic particles converge eventually to the Hubble flow \cite{Peebles_1993}. The conformally transformed tetra-velocity is
\begin{equation}
\accentset{\circ}{\mathbf{v}}= \gamma (\: \accentset{\circ}{\partial}_t + v^{(i)}\: \accentset{\circ}{\partial}_{(i)}),
\end{equation}
but now $\accentset{\circ}{\mathbf{v}}$ is not longer geodesic since $v^{(i)}$ is time-dependent. This means that the electromagnetic field solution of a geodesic (free-falling) charge not following the Hubble flow is equal to an accelerated charge solution in Minkowski space-time, generating a radiation field in the cosmological model we are considering. For small peculiar velocities, the velocity is $\accentset{\circ}{\vec{v}}= \accentset{\circ}{\vec{v}}_0/\Omega(t)$. The acceleration in the conformally flat space is then co-linear to the velocity, $\accentset{\circ}{\vec{a}}= - \accentset{\circ}{\vec{v}} H(t)$. The electric radiation field in the FLRW space-time can be written using the result in flat-space time as:
\begin{equation}
\vec{E}_\text{rad}=\frac{\accentset{\circ}{\vec{E}}_\text{rad}}{\Omega^2}= \frac{Q H(t_F)}{r} \Big[\vec{v}-(\vec{n}\cdot\vec{v}) \vec{n}\Big],
\end{equation}
where $\vec{v}$ is the peculiar velocity in the cosmic frame and $\vec{n}$ is unit director vector. The radiation rate, as measured by the Hubble flow frame, can be obtained from the Larmor formula in the conformal space. Transforming to the cosmic inertial frame, this radiation rate results:
\begin{equation*}
R=\frac{dE}{dt_F}
\end{equation*}
\begin{equation}
= \frac{dE}{dt} \Omega^{-1}= \frac{2}{3} Q^2(\vec{a}\cdot \vec{a}) \Omega^{-1}= \frac{2}{3} Q^2 \vec{v}_0^2 H(t_F)^2 \Omega^{-1},
\end{equation}
which goes to zero once the charge reaches the Hubble flow. For instance, if the charge starts with a peculiar velocity $v_0$ at $t_0$ in a universe described by $\Omega(t_F)= \mathbb{C} t_F^{\alpha}$ 
\begin{equation}
E=\int_{t_0}^{\infty} Rdt_F = \frac{2}{3} Q^2 \frac{\vec{v}_0^2 t_0^{-(1+\alpha)}}{\mathbb{C}}\frac{\alpha^2}{1+\alpha}.
\end{equation}

If the charge emits radiation, a self-force reaction appears, affecting the geodesic movement of the charge. In curved space-time, this self-force was found by DeWitt-Brehme and Hobbs (see an updated review in Ref. \cite{Poisson_Pound_Vega}). In a CFS, the non-local contribution to the self-force, the so-called tail-term, vanishes and thus the self-interaction is only local, given by:
\begin{equation}
v^{\mu} \nabla_{\mu} v^{\nu} = \frac{Q^2}{3m} (\delta^{\nu}_{\mu} +v^{\nu} v_{\mu}) R^{\mu}_{\; \rho} v^{\rho},
\end{equation}  
where the Ricci term expresses the local interaction between the electromagnetic field and the gravitational field. For small velocities, in the cosmic frame, we have:
\begin{equation}
\frac{d\vec{v}}{dt_F} =  H(t_F) \vec{v}-2\frac{Q^2}{m}\dot{H}(t_F)\vec{v}.
\end{equation}

When the particle moves comoving with the Hubble flow, $\vec{v}(t_0)=0$, there is no radiation as we have seen and, consistently, there is no self-force since the right hand side vanishes. If $\vec{v}(t_0)=\vec{v}_0 \neq 0$, the charge emits radiation, and choosing a model of universe again as $\Omega(t_F)= \mathbb{C} t_F^{\alpha}$, we can integrate the equations directly obtaining:
\begin{equation}
\vec{v}(t_F)= \frac{\vec{v}_0}{t^{\alpha}} \exp{\Big(-\frac{2 Q^2\alpha}{3m t_F}\Big)}.
\end{equation}

% Addition

The charged particle decays to the Hubble flow more rapidly than an uncharged particle, as seen in the cosmic frame, if $\alpha$ is positive. This result contradicts Haas et al. \cite{Haas_Poisson_2005}, who derive their result using conformal coordinates, integrating the equations within a small parameter approximation. If we integrate in \cite{Haas_Poisson_2005} the equation (6.2) exactly, we observe that the charged particle decays to the Hubble flow more rapidly than an uncharged particle when $\alpha'$ (as defined in a conformal scale factor, $a(\eta)=\eta^{\alpha'}$) is positive.

\subsection{Electromagnetic fields in spatially curved universes}

% I have changed the motivation of this section to avoid confusions.

 Electromagnetic fields in a spatially-curved universe have novel features with respect to the spatially-flat case. This is because the map between Minkowski and the FLRW metric is not global. The presence of spatial curvature, for instance,  could  modify  the  adiabatic  decay, as seen in the cosmic frame, on lengths close to the associated curvature scale, and thus explain some observational features of cosmic magnetic fields without introducing new physics \cite{Barrow_Tsagas_2008}.  We shall show now that there is a modification on the spatial dependence of the fields in the FLRW models.

Let us consider now a general FLRW metric:
\begin{equation}
ds^2= -dt +a(t) \Big[ \frac{dr^2}{1-Kr^2} +r^2 d\Omega^2 \Big].
\end{equation}

This metric has a null Weyl tensor for all values of $K$, which implies that all FLRW metrics are conformally flat. However, if $K\neq0$, then transformation to conformal coordinates is not global in general and it depends on the conformal time $T$ as well as the radial distance $R$,
\begin{equation}
ds^2= \Omega(T,R)^2 ds_0^2,
\label{eq: nonflat}
\end{equation}
where $a(t(T,R))\neq\Omega(T,R)$ and
\begin{equation}
ds_0^2:=-dT^2 + dX^2+dY^2+dZ^2.
\label{eq: minkosk}
\end{equation} 

In general, for $K \neq 0$, if we take an inertial frame $\accentset{\circ}{\mathbf{e}}_{a}$ in Minkowski space-time, the transformed conformal frame $\mathbf{e}_{a}= \accentset{\circ}{\mathbf{e}}_{a}/\Omega(T,R)$ is not inertial as follows from (\ref{eq: accelframe}). In turn, if $^F \mathbf{e}_a$ is a cosmic (inertial) observer adapted to (\ref{eq: nonflat}), then the corresponding $\accentset{\circ}{\mathbf{e}}_{a}$ flat frame will not be inertial in general. The electric and magnetic field in a cosmic frame are:
\begin{equation}
^F E^{(i)}=\: \accentset{\circ}{E}^{(i)} \Omega(T,R)^{-2}.
\label{eq: knonzeroe}
\end{equation}
\begin{equation}
^F B^{(i)}=\: \accentset{\circ}{B}^{(i)} \Omega(T,R)^{-2}.
\end{equation}
Note that this means that electric and magnetic fields, as seen by the frame comoving with the cosmic fluid, are related with a factor $\sim\Omega(T,R)^{-2}$ to Minkowskian electric and magnetic fields measured by a non-inertial frame $\accentset{\circ}{\mathbf{e}}_{a}$.  Moreover, it was shown in Ref. \cite{Gron_Johannesen_2011} that the transformation from $(t,r)$ to $(T,R)$ is not unique; so depending on the choice of $\Omega(T,R)$ we would have different Minkowskian frames. In the end, however, the combination would give unique $^F E^{(i)}$ and $^F B^{(i)}$. 

Let us take first an open universe, with $K=-1$. The line element can be written as:
\begin{equation}
ds^2= -dt +a(t)^2\Big[ d \chi ^2 + \sinh(\chi)^2 d\Omega^2 \Big].
\end{equation}

Defining $d\eta= dt/a(t)$, we obtain
\begin{equation}
ds^2= a(\eta)^2 \Big[-d\eta^2+ d \chi ^2 + \sinh(\chi)^2 d\Omega^2 \Big].
\label{eq: flwropen}
\end{equation}

The component of the inertial frame in these coordinates can be written as:
\begin{equation}
^F \mathbf{e}_{(0)}=\partial_{\eta}/a(\eta), \quad ^F \mathbf{e}_{(i)}= \partial_{\mu'}/\parallel \partial_{\mu'} \parallel.
\end{equation}

As it is shown in Ref.\cite{Gron_Johannesen_2011}, we can find a coordinate transformation from (\ref{eq: flwropen}) to (\ref{eq: minkosk}) with  of the form:
\begin{equation}
T= \frac{1}{2}[ f(\eta+\chi)+f(\eta-\chi)], \quad  R= \frac{1}{2}[ f(\eta+\chi)-f(\eta-\chi)],
\end{equation}
with the function $f$ given by a family of transformation
\begin{equation}
f(x):= c \Big[ b + \coth  \Big( \frac{x-a}{2} \Big)\Big] ^{-1} +d,
\end{equation}
and where
\begin{equation}
\Omega(T,R) = \frac{a(\eta) \sinh(\chi) }{R}.
\end{equation}

Choosing $f(x)=C e^{x}$ where $C$ is a constant, we have the transformation:
\begin{equation}
T=C e^{\eta} \cosh(\chi), \quad R =C e^{\eta} \sinh(\chi).
\end{equation}

For concreteness, let us take the radial component of an electric field, that would transform as:
\begin{equation}
^F E_{(r)} = F_{\mu \nu} e^{\mu}_{(r)} e^{\nu}_{(0)}
\end{equation}
so we have
\begin{equation*}
^F E_{(r)}= a(t)^{-2} F_{\mu \nu} (\partial_{r})^{\mu}(\partial_{\eta})^{\nu}=a(t)^{-2} \accentset{\circ}{F}_{AB} (\partial_{r})^{A}(\partial_{\eta})^{B}
\end{equation*}
\begin{equation}
=a(t)^{-2} \accentset{\circ}{F}_{AB} \frac{\partial \chi}{\partial r}(\partial_{\chi})^{A}(\partial_{\eta})^{B}=\frac{C^2 e^{2 \eta}}{a(t)^2 \sqrt{1+r^2}} \accentset{\circ}{E}_{(R)},
\end{equation}
where $ \accentset{\circ}{E}_{(R)}$ is a radial field as measured by a frame in the Minkowski space-time with coordinate $(T,R)$. From this expression, we can already note that for short distances to the center of the radial field, the space curvature is negligible. For instance, in the case of a Coulomb field, $ \accentset{\circ}{E}_{(R)}= Q/R^2= Q/(C e^{\eta} \sinh(\chi))^2$, we obtain:
\begin{equation}
^F E_{(r)} = \frac{Q}{a(t)^2 r^2 \sqrt{1+r^2}},
\end{equation}
or in radial coordinates $\chi$:
\begin{equation}
^F E_{(\chi)} = \frac{Q}{a(t)^2 \sinh(\chi)^2}.
\end{equation}

Note that the spatial curvature makes the field decay more rapidly than in flat space-time. Indeed, expanding for short distances we see that:
\begin{equation}
^F E_{(\chi)} = \frac{1}{a(t)^2} \Big( \frac{Q}{\chi^2} - \frac{Q}{3} +...\Big),
\end{equation}
where the field is weaker than the Coulomb field by a factor $\frac{Q}{3}$, where the $3$ comes from the number of spatial dimensions. This shows a very simple example on how the spatial curvature could affect the electromagnetic field.

Finally, let us briefly discuss the case of a closed universe, with $K=1$. Even though Maxwell's equations are conformally invariant, the change of topology in these universes forces to change the boundary conditions of the solution \cite{Infeld_Schild_1945}. For instance, the Coulomb field of a single charge is not a conformal solution in this space-time: from the conserved current, the total charge of a closed universe must be zero. We can show this re writing the FLRW metric for a closed universe as:
\begin{equation}
ds^2= a(\eta)^2 \Big[-d\eta^2+ d \chi ^2 + \sin(\chi)^2 d\Omega^2 \Big].
\end{equation} 

Solving Maxwell equations in these coordinates, Infeld and Schild have shown that the monopole solution is:
\begin{equation}
^F E_{(\chi)}= \frac{Q}{\sin(\chi)^2}.
\end{equation}

This solution has two singularities in $\chi=0$ and $\chi=\pi$ and the total charge is zero, as can be seen from Gauss law. Changing the origin of the frame to the center of the other singularity, $\chi'=\pi -\chi$, we have:
\begin{equation}
^F E_{(\chi')}= -\frac{Q}{\sin(\chi')^2}.
\end{equation}

We can interpret this solution as two opposite charges resting in the antipodes of the universe; the field lines start at $\chi=0$ and end in a negative image charge at $\chi=\pi$. In this case, the positive spatial curvature enhances the flat Coulomb field by a factor $Q/3$.
\begin{equation}
^F E_{(\chi)} = \frac{1}{a(t)^2} \Big( \frac{Q}{\chi^2} + \frac{Q}{3} +...\Big)  	
\end{equation}

The behavior of all these Coulomb fields combined are shown in Figure 1.

\begin{figure}
\begin{center}
\includegraphics[width=1\linewidth]{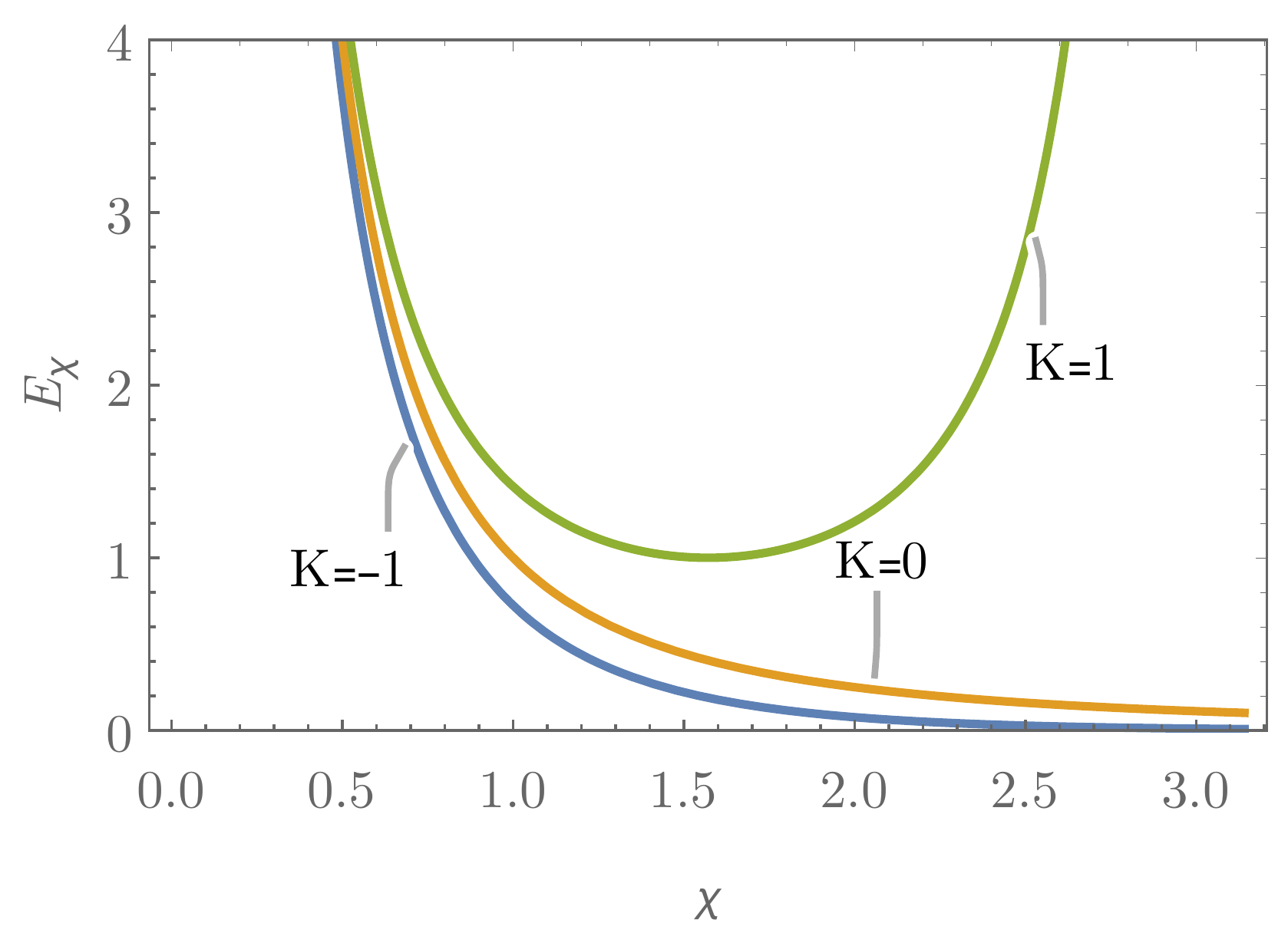}
\caption{Electric field of an static charge for different curvatures at a fixed time}
\end{center}
\end{figure}

\section{Cosmological effects on local electromagnetic fields}

Observers following the Hubble flow, as were characterized above, are free-falling in the FRLW universe. The spatial hyper-surfaces of constant time $t_F$ in the comoving coordinates (\ref{eq: flrw}) of this frame, have geodesic distances given by $a(t_F)r_F$. This means that the radial coordinate $r_F$ is not a true spatially geodesic coordinate, i.e. space as defined by this foliation is expanding \footnote{This can be shown in a coordinate independent way as in Ref. \cite{klein2011fermi}}. 
If we want to characterize local measurements within this space-time, we need a geodesic frame whose local description reduces to Minkowski near the frame and thus does not expand for close distances. This particular frame is the so-called local Fermi frame (LFF) \cite{Nesterov_1999}.
%Addition
This coordinate system has been recently used to obtain convenient expressions for cosmological observables \cite{Baldauf_Seljak_Senatore_Zaldarriaga_2011}. In the next section, we shall calculate the values of the electromagnetic field in this frame and use it to obtain the local radiation rate.

\subsection{Fermi coordinates in cosmology}

We shall apply the LFF to analyze the behavior of the EM field as measured by a local reference system, where the universe expansion appears as second-order corrections to the dynamics in Minkowski space-time (see Refs. \cite{Cooperstock_Faraoni_Vollick_1998} and \cite{Mashhoon_Mobed_Singh_2007}). These frames are constructed around an observer who follows a geodesic world-line $\gamma$ and carries an orthonormal tetrad $e_{a}(\tau)$, parallel transported through $\gamma(\tau)$. We can introduce a set of coordinates $\lbrace t_{L}, x_{L}^{i} \rbrace$ such that in a neighborhood of $\gamma(\tau)$, the metric assumes the form:
\begin{equation}
^L g_{00}= -1 - \: ^L R_{0i0j} (t_{L}) x_{L}^{i} x_{L}^{j}+\mathcal{O}(x_{L}^3),
\end{equation}
\begin{equation*}
^L g_{0i}= -\frac{2}{3} \: ^L R_{0ijk} (t_{L}) x_{L}^{j} x_{L}^{k}+\mathcal{O}(x_{L}^3),
\end{equation*}
\begin{equation*}
^L g_{ij}= \delta_{ij}-\frac{1}{3} \: ^L R_{ijkl} (t_{L}) x_{L}^{k} x_{L}^{l}+\mathcal{O}(x_{L}^3),
\end{equation*}
where we define:
\begin{equation}
\: ^L R_{\alpha \beta \gamma \delta}:= R_{\mu \nu \rho \sigma } \:  e^{\mu}_{\; (\alpha)}  e^{\nu}_{\; (\beta)}   e^{\rho}_{\; (\gamma)}  e^{\sigma}_{\; (\delta)}.
\end{equation}

At $\mathbf{x}_L=0$, the metric reduces to the Minkowski metric. These coordinates are valid in a tube-like region around the inertial path $\gamma$ with $r_L< \mathcal{L}$, where $\mathcal{L}$ is a measure of the radius of curvature of space-time. In order to obtain the Fermi coordinates of an expanding universe, we use FLRW coordinates as in (\ref{eq: flrw}). It is possible to construct Fermi coordinates around a fundamental inertial observer with proper time $\tau= t_{F}$ and associated frame $e^{\mu_{F}}_{(0)}= \delta^{\mu_{F}}_{(0)}$ and  $e^{\mu_{F}}_{(i)}= \Omega^{-1} \delta^{\mu_{F}}_{(i)}$ with the transformation up to third order given by:
\begin{equation*}
t_{F}=t_{L}-\frac{H(t_L)}{2} r_{L}^2+... ,
\end{equation*}
\begin{equation}
x^{i}_{F}=\frac{x^i_{L}}{\Omega(t_{L})} \Big( 1+ \frac{H(t_{L})^2}{4} r_{L}^2 \Big)+... ,
\label{eq: fermicoord}
\end{equation}
where $r_{L}^2:=\delta_{ij}x_L^i x_L^j$ and $H(t_L):= \dot{\Omega}/\Omega$ (see Ref. \cite{Baldauf_Seljak_Senatore_Zaldarriaga_2011}). In these coordinates, the metric up to second order hold:
\begin{equation}
ds^2= -\Big[ 1-\Big ( \dot{H} +H^2 \Big )r_L^2 \Big] dt_{L}^2 + \Big[1-\frac{1}{2} H^2 r_{L}^2 \Big] d^3x_{L}+...
\end{equation}

%Modification
Contrary to the comoving case, the integral paths  of spatial Fermi coordinates, $\vec{x}_L=$constant, are \textbf{not} geodesics. In Fermi coordinates, at first order, the geodesic deviation equation is given by:
\begin{equation}
\ddot{\vec{x}}_L-(\ddot{a}/a) \vec{x}_L=0,
\label{eq: geodesic}
\end{equation}
By choosing these coordinates, in the small velocity limit, we replace the expansion of space (as seen in the comoving frame) to a modification of the inertial structure: coordinates are fixed, and free-falling bodies are accelerated. The LFF, stationary in these coordinates can be described, up to second order, with the tetrad frame:
\begin{equation}
^L e_{a}^{\; \mu}= \delta_{a}^{\mu} + \psi_{a}^{\mu} \: r_{L}^2,
\end{equation}
where $\psi_{a}^{\mu}=$diag$\Big(-\frac{1}{2} (H^2+\dot{H}^2), \frac{1}{4} H^2,\frac{1}{4} H^2,\frac{1}{4} H^2 \Big )$, in Fermi coordinates. The kinematics of this frame, at the lowest order, is characterized by a radial acceleration given by
\begin{equation}
^L a^{i}= x^i_L \ddot{\Omega}(t_L)/\Omega(t_L),
\end{equation}
which depends on the acceleration of the universe and increases as we move away from the central inertial observer. This means that particles in rest with respect of this frame are accelerated, dragged by the acceleration of the universe. Conversely to the cosmic frame, the LFF is a rigid coordinate system at first order since expansion appears at the next order, $\Theta=-\frac{3}{2} H(t_L)\dot{H}(t_L)r^2_L$. From this expression, we see that if $\dot{H}\equiv0$, the frame is rigid, as in the case of a de Sitter universe.

\subsection{Electric and magnetic fields in the local frame}
Maxwell's equations in the LFF, at the lowest order expansion, read:
\begin{equation}
\dot{E}_{(i)}=  qH(t_L)^2(\vec{x_L} \times \vec{E})_{(i)} + (\vec{\nabla}\times \vec{B})_{(i)} -  j_{(i)},
\label{eq: locale}
\end{equation}
\begin{equation}
\dot{B}_{(i)}=  -qH(t_L)^2(\vec{x_L} \times \vec{B})_{(i)} - (\vec{\nabla}\times \vec{E})_{(i)},
\label{eq: localb}
\end{equation}
and 
\begin{equation}
\vec{\nabla} \cdot \vec{E} = \rho_{e}, \quad \vec{\nabla} \cdot \vec{B}= 0.
\end{equation}

where $q:=-\ddot{a}(t_L)a(t_L)/\dot{a}(t_L)^2$ is the deceleration parameter. These equations show that electromagnetic fields in the LFF are very different from the ones measured by the cosmic frame. 
Consider the case of a highly conducting medium, where, following Ohms law, we have $E_{(i)}=j_{(i)}=0$. The propagation equation for the magnetic field is:
\begin{equation}
\dot{B}_{(i)}=  qH(t_L)^2(\vec{x_L} \times \vec{B})_{(i)},
\label{eq: magneticlff}
\end{equation}
in contrast with the propagation as seen by the cosmic frame:
\begin{equation}
^F \dot{B}_{(i)}= -2H(t_F)\:^FB_{(i)}.
\end{equation}

Note that (\ref{eq: magneticlff}) depends on the acceleration state of universe and not on the expansion rate given by $H$. In a similar fashion, the dynamics of free-falling particles in the Newtonian approximation of the FLRW space-time are affected by a cosmological force that depends on the acceleration of the universe  \cite{Carrera_Giulini_2010}. For small time scales we can assume $q H(t_L)\sim q_0 H_0$, and then solve the propagation equation with the differential operator
\begin{equation}
\hat{\mathcal{D}}:=\exp{[t_L (q_0H_0) \vec{x}_{L} \times]}:= \sum^{\infty}_{n=0} (t_Lq_0H_0)^n(\vec{x}_L \times)^n/n!,
\end{equation}
that gives the Rodrigues's formula of a vector \cite{rodrigues}. With an initial field seed $\vec{B}_0$, the solution of (\ref{eq: magneticlff}) is:
\begin{equation}
\vec{B}(t)= \hat{\mathcal{D}} \vec{B}_0 \sim \vec{B}_0+ (q_0 H_0) t_L \vec{x}_{L} \times \vec{B}_0 + \mathcal{O}(\vec{x}_L^2),
\end{equation}
where the magnitude of the field is frozen in time $|\vec{B}(t_L)|=|\vec{B}_0|$. This solution could be applied to local cosmological scenarios, as galaxies and galaxy clusters, and will be explored elsewhere.

If the universe is accelerating, two systems following the Hubble flow have geodesic deviation that can be measured in the LFF using equation (\ref{eq: geodesic}). If one of these systems is charged, there is a relative acceleration that will induce, in principle, an electric radiation field as measured in the proper frame of the other system. In order to analyze this, we have to evaluate the fields in the LFF.
%%Addition$$$$
Note that the value of $E$ and $B$ on LFF and the cosmic frame will coincide on the geodesic $x_L=0$. However, building an extended frame is needed to evaluate quasi-local quantities such as the radiation rate. On the other hand, we need to specify a coordinate system to map the values of the fields to physical points in space-time.
%%%%%

In the next section, we investigate the simple case of a charged particle in a spatially flat universe as seen by the LFF.

\subsection{Charged particle and local radiation}

Let us consider a charge following the Hubble flow in a FLRW space-time. Choosing conformal coordinates, the particle is represented with the four-current:
\begin{equation}
j^{\mu}= Q \int d\tau \: \frac{dz^{\mu}}{d \tau} \: \frac{\delta^{(4)} (x^{\mu}-z^{\mu})}{\sqrt{-g}} \equiv  Q \: \delta^{(3)}(x^{i}) \: \delta^{\mu}_{0} \: \Omega^{-4},
\end{equation} 
where the particle is placed at the origin of the coordinate system $(x^{\mu}=0)$ and has a velocity given by $dz^{\mu}/d\tau=\delta^{\mu}_0/\Omega$ (see also the discussion in Ref. \cite{akhmedov2010classical}). As we have seen in the previous sections, the solution is the Coulomb field:
\begin{equation}
F_{i0}= \frac{Q}{r^3} \: x^i.
\label{eq: coulombf}
\end{equation}

The electric field measured by a cosmic observer is then given by $^F E^{(i)}= Q/(r_F^3 \Omega^2)\: x^i_F$, in accordance with (\ref{eq: emcosm}). Note that no magnetic field arises even though the electric field is now time-dependent. This can be justified from equations (\ref{eq: locale}) and (\ref{eq: localb}), showing that a time dependent solution of $E$ with $B=0$ is possible if the congruence is expanding. 

Now, we obtain the electromagnetic field of a charged particle as seen by a LFF. In order to do this, we first express the Coulomb field (\ref{eq: coulombf}) ---in CF coordinates--- into FLRW coordinates
\begin{equation}
^F F_{i0}= \frac{Q}{\Omega(t_{F}) r_F^3} \: x_F^i.
\end{equation}

The transformation matrix between FLRW coordinates and Fermi coordinates can be calculated from equations (\ref{eq: fermicoord}). In the new coordinates, the non-null components of Faraday's tensor are:

\begin{equation*}
^{L} F_{0i}=  \frac{\partial x_{F}^{\mu}}{\partial x_{L}^{0}} \frac{\partial x_{F}^{\nu}}{\partial x_{L}^{i}}\: ^{F} F_{\mu \nu}=
\end{equation*}
\begin{equation}
=Q \frac{x_L^i}{r_L^3}  - \frac{Q}{2} \frac{x_L^i}{r_L} \Big(H(t_L)^2- H(t_L)\frac{\ddot{a}}{a} \Big) + \mathcal{O}(Hr)^4,
\end{equation}

Then, there is no magnetic field in this frame and the electric field measured by these observers is given by:
\begin{equation*}
^L E^{(i)}= \: ^L F_{\mu \nu} \: ^L e^{\mu}_{(0)} \: ^L e^{\nu}_{(i)}
\end{equation*}
\begin{equation}
= Q \frac{x_L^i}{r_L^3}\Big[1+ H(t_L)^2 r_L^2 (1+2q) + \mathcal{O}(Hr)^4 \Big].
\label{eq: fermicoulomb}
\end{equation}

The first order term in (\ref{eq: fermicoulomb}) is the Coulomb field as expected from small $r_{L}$. The next correction to the Coulomb field depends on the square of the cosmic velocity $Hr$ and the state of acceleration of the universe. For instance, if we take a de Sitter universe, $1+2q\equiv -1$, then the universe accelerated expansion reduces the strength of the field and $^L E^{(i)}$ goes to zero as we approach the cosmological horizon $H^2r^2 \rightarrow 1$.

Note that in the case of the cosmic frame, the electric field decays with time (if the universe is expanding) and decays away from the charge as a Coulomb field. This difference arises, of course, from the different natures of the frames. In the cosmic frame, observers are inertial but expanding at all scales, inducing an adiabatic decay of the fields. In the LFF, the congruence is accelerating away from the inertial observer placed at $x_L=0$ and expanding at the next order. 

When the charge is comoving with the LFF, since the magnetic field is zero by symmetry, there is no energy flux, $\mathcal{Q}^{\mu}=0$. If the charge is not comoving with the geodesic observer but is placed near at $x^{\mu}=X \delta^{\mu}_1$ (in FLRW/conformal coordinates) then the charge is accelerating away as seen by the geodesic observer. Indeed, from (\ref{eq: fermicoord}), the position of the charge at first order in the LFF is $x^{\mu_L}=X_L \delta^{\mu_L}_1$, with $X_L=\Omega(t_L) X$. This means that the charge has a velocity given by
\begin{equation}
\vec{V}=d\vec{X}_L/dt_L = H(t_L) \vec{X}_L,
\end{equation}
where $\vec{X}_L:= X_L \partial_{1_L}$, and an acceleration given by $\vec{A}=d\vec{V}/dt_L= (\ddot{\Omega}/\Omega)\vec{X}^L$, as seen by the LFF. The Coulomb solution of a charge placed outside the origin in conformal coordinates is
\begin{equation}
F_{i0}= \frac{Q}{\bar{r}^3} \: (x^i-X\delta^{i}_{1}),
\end{equation}
where $\bar{r}=|x^i-X\delta^{i}_{1}|$. Proceeding as above, we obtain at second order in $H(t_L)$ and first order in $X_L$ a non-zero magnetic field given by:
\begin{equation}
\vec{B}= (\vec{V} \times \vec{E}) \Big[ 1 - \frac{1}{2}qH(t_L)^2r_L^2 + \mathcal{O}(Hr)^4 \Big ],
\end{equation}
where $\vec{E}$ is the Coulomb field. Note that the magnetic field is zero if the position of the charge is the origin or if the universe is not expanding, i.e. if $\vec{V}=0$; if the universe is not accelerated, $q=0$ and the magnetic field is analog to one measured by a boosted frame in flat space-time. This implies a non-zero Poynting flux given by
\begin{equation}
\vec{\mathcal{Q}} \sim \frac{Q^2 H(t_L) X_{L} \sin(\theta)}{4 \pi} \Big[ \frac{1}{r_L^4} +\frac{qH(t_L)^2}{2r_L^2} \Big] \breve{\theta}, 
\end{equation}
with magnitude
\begin{equation}
\mathcal{Q} \sim \frac{Q^2 H(t_L) X_{L}}{4 \pi} |\sin(\theta)| \Big| \frac{1}{r_L^4} + \frac{qH(t_L)^2}{2r_L^2} \Big |,
\end{equation}
where we have used spherical coordinates around the observer, with $\breve{\theta}$ the azimuth unit vector. Along the radial direction of the moving charge, there is no radiation as in the analog situation in flat space-time. Note that the $r^{-2}$ term is analog to a radiation type flux, driven by the acceleration of the universe; if the universe is not accelerated, then the Poynting flux goes like $r^{-4}$ as a uniform moving charge. On the other hand, is not entirely similar to a uniformly accelerated charge in Minkowski space-time because the $\sin(\theta)$ part and acceleration dependence is linear and not quadratic. 
\begin{figure}
\begin{center}
\includegraphics[width=1\linewidth]{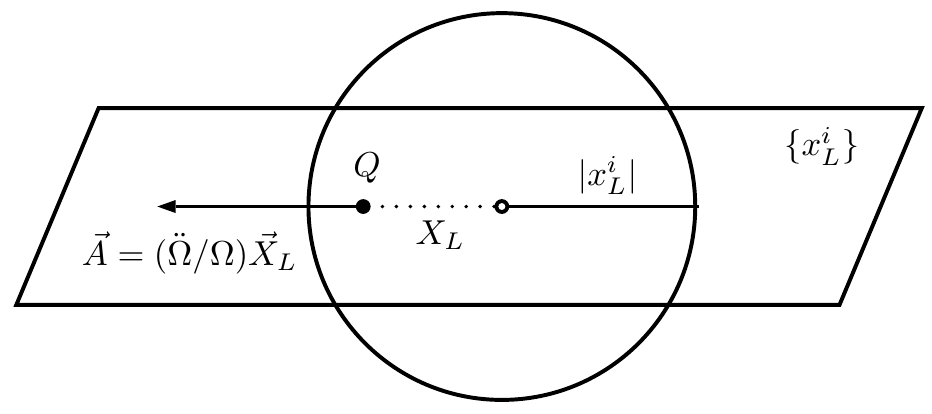}
\caption{Charged particle $Q$ accelerating in the local Fermi frame enclosed in a rigid sphere}
\end{center}
\end{figure}
Let us consider a rigid sphere around the observer that includes the accelerating charge. This rigid---not expanding--- sphere can be constructed in a FLRW with Fermi coordinates fixing a radius $|x^i_L|$, see Figure 2. Through this sphere, we can compute the radiation rate obtaining an analogue Larmor formula, given by:
\begin{equation}
R= \frac{1}{2} Q^2 A H(t_L).
\end{equation}

From the local reference frame, if the universe is expanding at an accelerated rate, the charged particle accelerates with the universe, and thus there is a non-null radiation. In this case, the energy provided as electromagnetic energy comes from the accelerated universe itself. By choosing a rigid sphere we compute fluxes without spurious deformation effects \cite{epp2009rigid}; however, in an expanding universe, this sphere requires energy to stay rigid, i.e. an external acceleration, that should be included in the final energy balance as gravitational energy. 

We stress that energy balance in curved space-time is a subtle subject. If the space-time has no time-like Killing vector, there is no obvious concept of conserved energy, i.e. energy is not a relativistic invariant in this case. This is associated with the problem of gravitational energy and its multiple definitions. An interesting unsolved problem is the formulation of a conceptual framework to treat, for instance, the conversion of gravitational energy into electromagnetic energy and vice-versa, see Ref. \cite{rigidframes} and Ref. \cite{combi2017gravitational}. for various approaches.

\section{Conclusions}

We have investigated the electromagnetic fields and charges in a cosmological background. We showed that although Maxwell's equations are conformally invariant, the well-known adiabatic decay only occurs in a frame comoving with the Hubble flow in the flat FLRW model. We proved that if a charged particle is free-falling but with a peculiar velocity, it has a radiation field, contrary to expectations. When the universe is spatially curved, we have shown that the field of a static charged changes its behavior depending on this curvature. In particular, if the universe is open, the field decays faster than the Coulomb field. To study the electromagnetic fields in local reference frames within an expanding space-time, we built a Fermi frame for a FLRW geometry. As seen by a local observer the electric and magnetic field presents novel interesting features. In particular, we analyzed the case of a charged particle. We have found that if the charge is accelerating with the cosmological expansion, the local frame detects non zero radiation.

\acknowledgments

This work was supported by the Argentine agency CONICET (PIP 2014-00338) and the Spanish Ministerio de Econom\'ia y Competitividad (MINECO/FEDER, UE) under grant AYA2016-76012-C3-1-P. We thank Federico Lopez Armengol for useful discussions.

\bibliographystyle{ieeetr}
\bibliography{bibliography}
\end{document}